\documentclass[12pt,a4paper]{article}
  \usepackage{a4wide}
  \usepackage{latexsym}
  \usepackage{epsf}
  \usepackage{amssymb}
  \usepackage{graphicx}
  \usepackage{amsmath, cite}
  \usepackage{amsmath,amssymb,amsthm}
  \usepackage{verbatim}
  \usepackage{hyperref}
\renewcommand{\d}{\textrm{d}}
\newcommand{\e}{\textrm{e}}
\newcommand{\ep}{\epsilon}
\newcommand{\w}{\wedge}

\newcommand{\be}{\begin{equation}}
\newcommand{\ee}{\end{equation}}
\newcommand{\ba}{\begin{eqnarray}}
\newcommand{\ea}{\end{eqnarray}}
\newcommand{\nn}{\nonumber}
\newcommand{\lp}{\left(}
\newcommand{\rp}{\right)}
\renewcommand{\k}{\kappa}

\begin{document}
\numberwithin{equation}{section}
\begin{flushright}
\small UUITP-28/10\\
\small ITP-UH-15/10\\
\small NSF-KITP-10-121\\

\date \\
\normalsize
\end{flushright}

\begin{center}

{\LARGE \bf{Smeared versus localised sources  \\
\vspace{0.3cm} in flux compactifications}} \\

\vspace{1 cm} {\large  Johan Bl{\aa}b\"ack$^\dagger$, Ulf
H.~Danielsson$^\dagger$, Daniel Junghans$^\ddagger$, \\
\vspace{0.2cm}Thomas Van Riet$^\dagger$, Timm Wrase$^\ddagger$$^\sharp$ and Marco Zagermann$^\ddagger$$^\natural$}\\

\vspace{.8 cm}  {${}^\dagger$ Institutionen f{\"o}r fysik och
astronomi\\
Uppsala Universitet, Box 803, SE-751 08 Uppsala, Sweden}\\
\vspace{0.2cm} {\upshape\ttfamily johan.blaback, ulf.danielsson,
thomas.vanriet@fysast.uu.se} \\

\vspace{0.4 cm}  {${}^\ddagger$ Institut f{\"u}r Theoretische Physik \&\\
Center for Quantum Engineering and Spacetime Research\\
Leibniz Universit{\"a}t Hannover, Appelstra{\ss}e 2, 30167
Hannover, Germany}\\

\vspace{0.2cm} {\upshape\ttfamily daniel.junghans,
marco.zagermann@itp.uni-hannover.de} \\

\vspace{0.4 cm}  {${}^\sharp$ Department of Physics, Cornell University, Ithaca, NY 14853, USA}\\

\vspace{0.2cm} {\upshape\ttfamily timm.wrase@cornell.edu} \\

\vspace{0.4 cm}  {${}^\natural$ Kavli Institute for Theoretical Physics, Santa Barbara, CA 93106, USA}\\

\vspace{1cm}

{\bf Abstract}
\end{center}

\begin{quotation}
We investigate whether vacuum solutions in flux compactifications
that are obtained with smeared sources (orientifolds or D-branes)
still survive when the sources are localised. This seems to rely on
whether the solutions are BPS or not. First we consider two sets of
BPS solutions that both relate to the GKP solution through
T-dualities: $(p+1)$-dimensional solutions from spacetime-filling
O$p$-planes with a conformally Ricci-flat internal space, and $p$-dimensional
solutions with O$p$-planes that wrap a 1-cycle inside an everywhere
negatively curved twisted torus. The relation between the solution
with smeared orientifolds and the localised version is worked out in
detail. We then demonstrate that a class of non-BPS AdS$_4$
solutions that exist for IASD fluxes and with smeared D3-branes
(or analogously for ISD fluxes with anti-D3-branes)
does not survive the localisation of the (anti) D3-branes. This casts
doubts on the stringy consistency of non-BPS solutions that
are obtained in the limit of smeared sources.
\end{quotation}

\newpage
\tableofcontents

\section{Introduction}
Many compactifications that give rise to vacua with
phenomenologically appealing properties feature spacetime-filling
sources such as orientifold planes or D-branes. In most cases these solutions are derived in the
limit that the sources are smeared, although some localised
solutions are known, see e.g.~\cite{Giddings:2001yu,Schulz:2004ub,
Grana:2006kf}. Smearing means that delta function sources in the
equations of motion are replaced with specific regular functions
that integrate to the same value.

Apart from simplifying the task of finding solutions, reductions
with smeared sources may allow for consistent truncations in certain
compactifications on group spaces or coset manifolds (see
e.g.~\cite{Angelantonj:2003rq, Derendinger:2004jn}). However, an
orientifold plane is defined through its involution and a D-brane
by its boundary conditions, which makes them
really localised objects. It is therefore important to study the corrections to a smeared solution
that arise upon localising an orientifold plane or a D-brane. A notable difference
between smeared and localised solutions is that the equations of
motion, in the localised case, necessarily imply that the spacetime
is warped (which can be appealing to solve the hierarchy problem
\cite{Randall:1999ee}).

One way to incorporate the changes that arise upon localising a source is
through ``warped effective field theory'' \cite{DeWolfe:2002nn,
Giddings:2005ff, Martucci:2006ij, Frey:2006wv, Koerber:2007xk, Douglas:2008jx,
Shiu:2008ry, Frey:2008xw, Marchesano:2008rg, Martucci:2009sf, Chen:2009zi,
Douglas:2009zn}, in which one derives the correction to the
four-dimensional effective action. Another way, which we are
pursuing in this paper, is to work directly with the ten-dimensional
equations of motion.

There are only few known solutions with localised sources, and we
therefore first generalise some existing solutions to other
spacetime dimensions. All the solutions turn out to be T-dual to
the known four-dimensional solution of \cite{Giddings:2001yu}. In
the smeared limit these solutions either have an internal space
that is Ricci-flat, being the $D$-dimensional generalisation of
the GKP solution \cite{Giddings:2001yu}, or the solutions have a
negatively curved twisted torus as internal space, being the
$D$-dimensional generalisation of some solutions given in
\cite{Kachru:2002sk} (which themselves are T-dual to the GKP
solution, see also \cite{Gurrieri:2002wz}). The solutions on the twisted tori of
\cite{Kachru:2002sk} were obtained in the smeared limit and a
discussion on their localisation was given in \cite{Schulz:2004ub,
Grana:2006kf}. Here we discuss the $D$-dimensional generalisation
of these solutions very explicitly from the point of view of the
10-dimensional equations of motion. Furthermore we do not assume
flux configurations that are necessarily supersymmetric, in
contrast to most references on solutions from twisted tori.
However, the solutions will be BPS, in the same sense that the GKP
solution is BPS but not necessarily supersymmetric. The phenomenon
of BPS but non-susy solutions has been given a higher-dimensional
interpretation in \cite{Lust:2008zd}, where the BPSness turns out
to be directly related to the existence of brane calibrations even
in absence of SUSY.

Apart from providing a larger playground for studying localisation
effects there is a general interest in constructing the
$D$-dimensional generalisation of the four-dimensional string
landscape. Partly because some stringy consistency issues might be
easier to deal with in lower dimensions (see e.g.~the recent
discussion in \cite{Banks:2010tj}) and partly because the landscape
is really featuring vacua of all kinds of dimensions and there might
be transitions between vacua of various dimensions (see
e.g.~\cite{Carroll:2009dn}).

Our motivation for this paper comes from the interesting observation
made in \cite{Douglas:2010rt} that string/M theory
compactifications, at tree-level, using manifolds whose curvature is
everywhere negative, must have significant warping. Since part of
our solutions, in the smeared limit, have negatively curved spaces
we can address this issue in concrete examples. The reason that
significant warping is required is that, for \emph{unwarped}
metrics, one can show that negative curvature in the internal space
requires a source of energy momentum with negative tension at every
point. Hence, imagine that in the localised case there exists a
regime in which warping can be neglected compared to the fluxes.
Then in that regime we require a source of negative tension at every
point in the internal space. This leads to a contradiction since
this regime is by definition far away from the delta-like
orientifold source, where warping is strong.  We elaborate further
on this argument and illustrate it with our simple examples.  Specifically, we want to emphasize the r\^ole of the BPS condition in
achieving localisation.  Furthermore the T-duality chain of the BPS
solutions shows that the argument of \cite{Douglas:2010rt} for large warping corrections to
solutions with negatively curved internal spaces extends also to
solutions with Ricci-flat internal spaces and $H$-flux. As an important application of our results, we  show that the properly integrated negative curvature of the smeared twisted tori solutions stays negative upon the localization of the O-planes, contrary to some naive expectations.

The main conclusion of our examples is that the individual
localisation corrections\footnote{Localisation corrections include,
apart from the warp factor, corrections due to the dilaton that
varies in the internal space and the non-zero field strength that is
sourced by the orientifold or D-brane.} cancel against each other
in the effective potential for certain BPS solutions (along the lines of \cite{DeWolfe:2002nn}).
However, there is in general no reason to expect an analogous cancelation of localisation
effects for non-BPS solutions. As an illustration, we explicitly
construct, in section \ref{nonBPS}, AdS solutions with smeared sources that are non-BPS by
going beyond the ISD flux configurations. Then we show how AdS$_4$ solutions
with D3-branes and imaginary anti-self-dual (IASD) fluxes (or analogously anti-D3-branes with ISD fluxes)
do \emph{not} survive localisation. We discuss the possible implications of our results
in section \ref{Discussion}.

\section{Type II supergravity}

To establish our notation and conventions, we present the equations
of motion for type IIA/B supergravity with O$p$-sources in Einstein
frame (D$p$-branes will be considered in section \ref{nonBPS}). We use the conventions of \cite{Koerber:2010bx} but go to Einstein frame and change the sign of $H$ (see also appendix A of \cite{Danielsson:2009ff}). Compared to \cite{Koerber:2010bx} our solutions with $\mu_p > 0$ correspond to O-planes for $p=2,3,6$ and anti-O-planes for $p=1,4,5$ and analogously for D-branes which have $\mu_p < 0$. Note that one can always flip the sign of all RR-fields, which leaves the closed string action invariant and maps O-planes/D-branes to anti-O-planes/anti-D-branes.
Throughout the paper $a,b$ are 10D indices; $\mu, \nu$ are along the orientifold plane and $i,j$ are
transverse to it. The common bosonic sector contains the metric $g_{ab}$, the dilaton
$\phi$ and the $H$ field strength. The RR sector of (massive) type IIA consist
of the $(F_0,)\, F_2, F_4$ field strengths, whereas in IIB one has the field
strengths $F_1, F_3, F_5$, with $F_5$ satisfying $F_5=\star F_5$.

The trace reversed Einstein equation is
\ba
R_{ab} &=& \tfrac{1}{2} \partial_a \phi \partial_b \phi + \e^{-\phi} \lp \tfrac{1}{2} |H|^2_{ab}-\tfrac{1}{8} g_{ab} |H|^2 \rp\\
&&+\sum_{n\leq 5} \e^{\tfrac{5-n}{2} \phi} \lp \tfrac{1}{2(1+
\delta_{n5})} |F_n|^2_{ab}-\tfrac{n-1}{16(1+ \delta_{n5})} g_{ab}
|F_n|^2 \rp +\tfrac{1}{2}(T^{loc}_{ab} -
\tfrac{1}{8}g_{ab}T^{loc})\,,\nn
\ea
where $\delta_{n5}$ is the Kronecker delta, and $|A|^2_{ab} \equiv \tfrac{1}{(p-1)!}\,A_{a a_2\ldots a_p}A_{b}^{\,\,a_2\ldots
a_p}$, $|A|^2 \equiv \tfrac{1}{p!}\,A_{a_1\ldots a_p}A^{a_1\ldots a_p}$.

The non-vanishing part of the local stress tensor is given by\footnote{Here and in the following, $\delta(Op)$ is meant to implicity also include sums of parallel O$p$-planes.}
\begin{equation}
T_{\mu\nu}^{loc}=\e^{\tfrac{p-3}{4}\phi} \mu_p g_{\mu\nu}
\delta(Op),\qquad \mu,\nu=0,1,\ldots, p\,,
\end{equation}
where $\mu_p$ is a positive number for an orientifold source and $\delta(Op)$ is a delta distribution with support on the O$p$-plane world volume. The dilaton equation of motion is given by
\be
\nabla^2  \phi= - \e^{-\phi} \tfrac{1}{2} |H|^2 +
\sum_{n\leq 5} \e^{\tfrac{5-n}{2} \phi} \tfrac{5-n}{4} |F_n|^2 -
\tfrac{p-3}{4} \e^{\tfrac{p-3}{4}\phi} \mu_p \delta(Op).
\ee
The Bianchi identities for the field strength are
\ba
\d H &=& 0,\\
\d F_n  &=& H \w F_{n-2} - \mu_{8-n} \delta_{n+1}(O(8-n)),\nn
\ea
where $\delta_{n+1}(O(8-n))$ is shorthand for the normalized $(n+1)$ volume
form transverse to the O$(8-n)$ orientifold plane multiplied by $\delta(O(8-n))$.
The equations of motion for the RR field strengths,
\be
\d \lp \e^{\tfrac{5-n}{2} \phi} \star F_n \rp = - \e^{\tfrac{3-n}{2} \phi} H \w \star F_{n+2} - (-1)^{\tfrac{n(n-1)}{2}} \mu_{n-2} \delta_{11-n}(O(n-2)),
\ee
can be obtained from the RR Bianchi identities for $n>5$ upon employing the rule $\e^{(5-n)\phi/2}F_n =(-1)^{(n-1)(n-2)/2}\,\star F_{10-n}\,$.

Finally, the equation of motion for the $H$ field strength is given by
\be
\d(\e^{-\phi}\star H)=-\tfrac{1}{2}\sum_{n}\e^{\frac{5-n}{2}\phi}\star F_n\w F_{n-2}\,,
\ee
where the sum over $n$ includes all even/odd numbers up to 10 for IIA/IIB.

\section{BPS solutions with Ricci-flat internal space}\label{MinkowskiRicciflat}
In this section, we consider a flux compactification down to $p+1$
dimensions with a spacetime-filling orientifold plane that has a
pointlike extension in the internal space (i.e.~an O$p$-plane). For $p=3$ this is
the famous GKP solution \cite{Giddings:2001yu}. The generalisations
we find here are written down for general $p = 1, \ldots, 6$.

\subsection{The smeared solutions}
When we look for solutions with smeared orientifold sources, we
assume that the dilaton, $\phi$, is constant and that the metric has
the form of a direct product
\begin{equation}
\phi=\phi_0\,,\qquad \d s^2_{10}=\d s_{p+1}^2 + \d s_{9-p}^2\,.
\end{equation}
The non-zero form fields are $H$ and $F_{6-p}$. The rest of the
RR-fields are identically zero.

The orientifold source enters the Einstein equation, the dilaton
equation (unless $p=3$) and the Bianchi identity for the $F_{8-p}$
field. Smearing implies that the delta form function in the Bianchi
identity is set equal to the normalized internal volume form, $\epsilon_{9-p}$,
and in the Einstein and dilaton equations the delta functions are set equal to one.
For a flat external space (Minkowski vacuum), this gives the
following, external Einstein equation (in form notation)
\begin{equation}
0 = -\tfrac{1}{8}\e^{-\phi_0}\star_{9-p}H\w H - \tfrac{5-p}{16}\e^{\tfrac{p-1}{2}\phi_0} \star_{9-p}F_{6-p}\w F_{6-p} + \tfrac{7-p}{16}\mu_p\,\e^{\tfrac{p-3}{4}\phi_0}\ep_{9-p}\,.\label{external}
\end{equation}
Since $F_{8-p}=0$, its Bianchi identity becomes
\begin{equation}\label{eq:F9bianchi}
0 = H \w F_{6-p} - \mu_p\,\ep_{9-p}\,.
\end{equation}
Combining \eqref{external} and \eqref{eq:F9bianchi} we can eliminate the source term to get
\begin{equation}
0 = H\w F_{6-p} - \tfrac{2}{7-p}\e^{-\tfrac{p+1}{4}\phi_0}
\star_{9-p}H\w H - \tfrac{5-p}{7-p} \e^{\tfrac{p+1}{4}\phi_0}
\star_{9-p} F_{6-p}\w F_{6-p}\,.
\end{equation}
To solve the above relation we apply the following Ansatz
\begin{equation}
F_{6-p} = (-1)^p \e^{-\tfrac{p+1}{4} \phi_0} \kappa \star_{9-p} H\,,
\end{equation}
which provides a second order equation in $\kappa$. To later be able
to solve the dilaton equation, it turns out that one solution has to
be discarded. The remaining solution is
\begin{equation}\label{eq:ISDsmear}
\boxed{F_{6-p} = (-1)^{p}\e^{-\tfrac{p+1}{4}\phi_0} \star_{9-p} H\,.}
\end{equation}
For the special case of $p=3$ this is the so-called ISD condition on
the $G$-flux \cite{Giddings:2001yu}. For $p=1$ there is a subtlety
because of the self-duality of $F_5$. The same derivation still
applies as long as one carries the self-duality around, and the
result is that (\ref{eq:ISDsmear}) needs to be adjusted by adding
the (10-dimensional) Hodge dual piece to the right hand side.

We will refer to the duality (\ref{eq:ISDsmear}) as the \emph{BPS
condition} for reasons that become clear below. Note, that just as in GKP,
the BPS condition does not necessarily imply supersymmetry. For constant dilaton
the BPS condition equates the $H$ equation of motion to the
$F_{6-p}$ Bianchi identity, and vice versa. Furthermore, using the
BPS condition in (\ref{eq:F9bianchi}) allows us to express the values
of the fluxes in terms of $\mu_p$,
\begin{equation}\label{eq:F9binachireduced} \mu_p =
\e^{-\tfrac{p+1}{4} \phi_0} |H|^2 = \e^{\tfrac{p+1}{4} \phi_0}
|F_{6-p}|^2\,.
\end{equation}
The dilaton equation of motion gives
\begin{equation}
0 = \nabla^2 \phi_0 =  -\tfrac{1}{2}\e^{- \phi_0} |H|^2 +
\tfrac{p-1}{4} \e^{\frac{p-1}{2}\phi_0} |F_{6-p}|^2
-\tfrac{p-3}{4}\e^{\tfrac{p-3}{4} \phi_0}\,\mu_p\,.
\end{equation}
The BPS relation \eqref{eq:ISDsmear} combines the $|F_{6-p}|^2$ term with the $|H|^2$
term, such that
\begin{equation}
0 = \tfrac{p-3}{4}\e^{- \phi_0} |H|^2
-\tfrac{p-3}{4}\e^{\tfrac{p-3}{4} \phi_0}\,\mu_{p}\,.
\end{equation}
This is solved trivially for $p=3$ and reduces to
(\ref{eq:F9binachireduced}) for $p\ne 3$.

The internal Einstein equation is
\begin{equation}
\begin{split}
R_{ij} &= \left(-\tfrac{1}{8}\e^{- \phi_0}|H|^2
-\tfrac{5-p}{16}\e^{\frac{p-1}{2} \phi_0} |F_{6-p}|^2 -
\tfrac{p+1}{16}\e^{\tfrac{p-3}{4} \phi_0}\,\mu_p\right)g_{ij}\\
&\quad + \tfrac12 \e^{- \phi_0} |H|^2_{ij}
+ \tfrac{1}{2}\e^{\frac{p-1}{2} \phi_0} |F_{6-p}|^2_{ij} \,,
\end{split}
\end{equation}
which by the use of equations \eqref{eq:ISDsmear}, \eqref{eq:F9binachireduced} tells us that the internal space is
Ricci-flat\footnote{We have used the BPS condition \eqref{eq:ISDsmear} to
rewrite $|F_{6-p}|^2_{ij} = \e^{-\tfrac{p+1}{2} \phi_0} \left( |H|^2 g_{ij} -
|H|^2_{ij}\right)$. This relation is not present in the case of
$p=6$, since $F_0$ has no indices.}
\begin{equation}
R_{ij}=0\,.
\end{equation}

\subsection*{Summary of the solution}
The non-zero fields in the Ansatz for the smeared orientifold
solution are (where $p=1,\ldots, 6$)
\begin{equation}
\phi = \phi_0\,,\quad H\,,\quad F_{6-p} \,,\quad R_{ij} \,.
\end{equation}
This leads to a $(p+1)$-dimensional Minkowski solution provided the
following conditions are satisfied
\ba
F_{6-p} &=& (-1)^p \e^{-\tfrac{p+1}{4}\phi_0} \star_{9-p} H \,, \label{eq:dual}\\
\d H &=& 0\,, \quad \d F_{6-p} = 0\,, \label{eq:closed} \\
\mu_p &=& \e^{-\tfrac{p+1}{4}\phi_0} |H|^2= \e^{\tfrac{p+1}{4} \phi_0} |F_{6-p}|^2\,,\label{charge}\\
R_{ij}&=&0\,.\label{flat}
\ea
For $p=1$ the Hodge dual piece needs to be added to the expression for $F_5$.

\subsection{The localised solutions}
We expect a localised orientifold to (i) induce a warping, (ii) source
the $F_{8-p}$ field strength and (iii) lead to a dilaton that varies in the
internal space. This can be seen from the standard D$p$-brane solutions
in asymptotically flat space. Therefore, if we consider a localised orientifold in
a flux background, the solution should allow two limits; a limit in
which the smeared flux background is found and a limit in which the
D$p$-brane solution in asymptotically flat space is found after
eliminating the background fluxes. In that sense one could view the
localised solutions, if one exists, as a superposition of two solutions.
So, as argued, the non-zero fields are
\begin{equation}
F_{8-p}\,,\ F_{6-p}\,,\ H\,, \phi\,.
\end{equation}

The metric Ansatz is given by
\begin{equation}\label{eq:metricAnsatz}
\d s_{10}^2=\e^{2a A}\d \tilde{s}_{p+1}^2 + \e^{2b A}\d
\tilde{s}_{9-p}^2\,,
\end{equation}
where $a$ and $b$ are some numbers to be determined later, and $A$ is
a function of the internal coordinates $x^i$ and is called the warp
factor. The external and internal metric are written as
\begin{equation}
\d \tilde{s}_{p+1}^2 =\tilde{g}_{\mu\nu}\d x^{\mu}\d
x^{\nu}\,,\qquad \d \tilde{s}_{9-p}^2 =\tilde{g}_{ij}\d x^{i}\d
x^{j}\,.
\end{equation}
So tildes are used when the warp factor dependence is taken out.
This will also apply to covariant derivatives ($\tilde{\nabla}$),
squares of tensors ($\tilde{T}^2$) and so on.

There is an ambiguity in what we call the internal metric
$\tilde{g}_{ij}$ since we can always absorb powers of $A$. Hence the
number $b$ can be seen as a gauge choice. Nonetheless, when one
considers how warping arises upon localising a source then the
internal metric has an absolute meaning as the internal metric
before localisation.

The Ricci tensor for the metric Ansatz \eqref{eq:metricAnsatz} reads
\begin{align}\label{eq:ricci}
R_{\mu\nu}&= \tilde{R}_{\mu\nu} - \e^{2(a-b)A} \tilde{g}_{\mu\nu}\Bigl(a [(p+1)a+(7-p)b](\tilde{\partial} A)^2 + a\tilde{\nabla}^2A \Bigr) \,, \nn \\
R_{ij}&= \tilde{R}_{ij} - \Bigl( b [(p+1)a+(7-p)b] (\tilde{\partial} A)^2+ b\tilde{\nabla}^2 A
\Bigr)\tilde{g}_{ij} \\
&+ \left[b\lp(p+1)a+(7-p)b\rp -(p+1)a (a-b) \right]\partial_i A\partial_j A
-[(p+1)a+(7-p)b]\tilde{\nabla}_i\partial_jA\,.\nonumber
\end{align}
There are no mixed components. The solution for the sources in flat
space have
\begin{equation}\label{a_and_b}
(p+1)a+(7-p)b=0\,,
\end{equation}
which is what we assume from now on. We furthermore choose the
normalisation of $A$ such that $a=1$. Note that this makes the
expression for the Ricci tensor much simpler.

The Ansatz for the $F_{8-p}$ field strength is
\begin{equation}\label{AnsatzF8-p}
F_{8-p}=-\e^{-2(p+1)A-\frac{p-3}{2}\phi}\tilde{\star}_{9-p}\d\alpha\,.
\end{equation}
For $p=3$, we  have to add the Hodge dual, and this coincides exactly with the
Ansatz of GKP \cite{Giddings:2001yu}. For $p=2$ there is also a
subtlety since $F_{8-p}$ is dual to $F_{6-p}$. Therefore, for $p=2$
the dual of the above expression for $F_6$ needs to be \emph{added}
to the expression for $F_4$ we will now construct (see \eqref{eq:F4F6}).

We obtain two equations by combining the Bianchi identity for
$F_{8-p}$ with the traced external Einstein equation to remove the
source, and by combining the dilaton equation with the traced external
Einstein equation to remove the source. After some algebra, these
two equations can be combined into
\begin{align}
\tilde{\nabla}^{2}\left(  \e^{(p+1)A+\frac{p-3}{4}\phi}+\left(
-1\right)
^{p}\alpha\right)    & =\e^{\frac{\left(p-3\right)  ^{2}}{p-7}A}\e^{\frac{p-3}{4}\phi}\tilde{R}_{p+1}\label{BPS}\\
& + \e^{\tfrac{(p+1)(9-p)}{p-7} A-\frac{p-3}{4}\phi}\left\vert \partial\left(
\e^{(p+1)A+\frac{p-3}{4}\phi}+\left(  -1\right)  ^{p}\alpha\right)  \right\vert ^{2}\nn\\
& +\tfrac{1}{2}\e^{\frac{(p+1)\left( p-5\right)
A}{p-7}+\frac{3p-5}{4}\phi }\left\vert
F_{6-p}-(-1)^{p}\e^{-\frac{p+1}{4}\phi}\star_{9-p}H\,\right\vert
^{2}\,,\nn
\end{align}
where the squares in the last two term are with respect to the warped metric. Since
the left hand side integrates to zero on a compact space, we find
that $\tilde{R}_{p+1}\leq 0$. For Minkowski solutions
($\tilde{R}_{p+1}=0$), both squares need to vanish, and we recover
the duality condition (\ref{eq:ISDsmear}) together with
\begin{equation}\label{alpha}
\alpha=(-1)^{p+1}\e^{(p+1)A+\tfrac{p-3}{4}\phi} + cst\,.
\end{equation}
Note that these results are obtained without the use of any other
equations than the traced external Einstein equation, the dilaton
equation and the Bianchi identity for $F_{8-p}$. Let us therefore
discuss how all other equations are solved. First of all we assume
the Bianchi identities $\d H=0$ and $\d F_{6-p}=0$. A rather lengthy
calculation\footnote{This computation goes along the same lines of
the smeared case but is more involved. Useful identities are the
expression for the Ricci tensor (\ref{eq:ricci}), the expression for
a Laplacian: $ \d\tilde{\star}_{9-p} \d \alpha =
(-1)^p\tilde{\nabla}^2\alpha \tilde{\star}_{9-p} 1$ and the relation
$\star_{10} (A_n\w B_m) = (-1)^{n(9-p-m)} \star_{p+1} A_n\w
\star_{9-p} B_m$, where $A_n$ is an external $n$-form and $B_m$ an
internal $m$-form.} then shows that all the other equations are
satisfied if the following conditions are met
\begin{align}
&\tilde{R}_{ij}=0\,,\label{conformalflat}\\
&\tilde{\nabla}^2 \left( \tfrac{4(p-3)}{7-p} A - \phi\right) =0
\qquad \Longrightarrow \qquad \phi=\tfrac{4(p-3)}{7-p} A + \phi_0\,,
\label{eq:phi2a} \\
&\tilde{\nabla}^2 \e^{\tfrac{16}{p-7}A}
= -\e^{-\phi_0}|\tilde{H}|^2 + \e^{\frac{p-3}{4}\phi_0}\mu_p \tilde{\delta}(Op)\,.\label{warpequation}
\end{align}
We used in the second equation that a regular harmonic
function on a compact space is constant and in the last equation
we have pulled out the warp factor in $|\tilde{H}|^2$ and $\tilde{\delta}(Op)$\footnote{The delta function is proportional to $1/\sqrt{g^{(9-p)}}$. This means that we have set the overall volume of the transverse space to one when we take the smeared limit $\delta(Op) \rightarrow 1$.}.

\subsection*{Summary of the solution}
The Minkowski solution obtained in the smeared limit allows a
localisation by adding a warp factor, $A$, in the metric Ansatz
\eqref{eq:metricAnsatz}, \eqref{a_and_b}. The varying dilaton can be
written in terms of the warp factor via (\ref{eq:phi2a}), and the
$F_{8-p}$ field strength can also be written in terms of the warp
factor via \eqref{AnsatzF8-p}, \eqref{alpha}. The value of the
warpfactor itself is then determined by the orientifold charge (and
related fluxes) through equation (\ref{warpequation}).

As announced earlier, when $p=2$, there is a subtlety since $F_4$ and $F_6$ are each others
dual and the solution needs some adjustment
\begin{equation}\label{eq:F4F6}
F_4= \e^{-\frac{3}{4}\phi}\star_{7}H
+\e^{-6A+\tfrac12 \phi}\star_{10}\tilde{\star}_{7}\,\d\alpha\,.
\end{equation}

If one uses the expression for the dilaton in terms of the warp factor
(\ref{eq:phi2a}), then one finds that the BPS equation
(\ref{eq:dual}) has not changed since
\begin{equation}
F_{6-p} = (-1)^p \e^{-\tfrac{p+1}{4}\phi} \star_{9-p} H = (-1)^p \e^{-\tfrac{p+1}{4}\phi_0} \tilde{\star}_{9-p} H\,,\label{eq:dualunchanged}
\end{equation}
where the first Hodge star is taken with respect to the warped metric.
Therefore the geometric moduli that are fixed by this BPS equation
\eqref{eq:ISDsmear} have not shifted position due to the warping. The
interpretation of this in terms of an effective potential $V_{eff}$
is that, at the Minkowski point, the contribution in $V_{eff}$
coming from the warped metric cancels the source contribution of
$F_{8-p}$ in $V_{eff}$ (see also \cite{DeWolfe:2002nn}). The integrated version of equation (\ref{warpequation})
also implies that (\ref{charge}) is still valid after
localisation. Furthermore, the condition \eqref{eq:closed} remains unchanged,
while the internal space changes from Ricci-flat
(\ref{flat}) to conformally Ricci-flat (\ref{conformalflat}).

Finally, we mention that for $p=6$ this solution is related to the
``massive D6 solution'' of \cite{Janssen:1999sa}, which considered
the same setup, but in a non-compact setting where the O6 is
replaced by a D6. This probably implies that a non-compact
version, for which the O$p$ source is replaced by a D$p$ source
exists for all values of $p$ we considered here, which generalises
some results in \cite{Janssen:1999sa}.

\section{BPS solutions with negatively curved twisted tori}

Consider the setup above with a smeared O$p$-plane whose tadpole is
canceled by $H$- and $F_{6-p}$-flux. A formal T-duality along one
direction of the $F_{6-p}$-flux or one direction along the O$p$-plane
maps this setup into a smeared O($p\pm1$)-plane whose tadpole
is canceled by $H$- and $F_{6\mp1-p}$-flux. Now we want to study
cases that arise after one T-duality along the direction of the $H$-flux.
If we start out with a torus as compact transverse space, then
this leads to twisted tori that have negative curvature. Therefore,
these setups are directly addressing the issue of
\cite{Douglas:2010rt}. There the authors show that compactifications
on spaces with negative curvature that lead to dS or Minkowski
solutions require a warp factor whose contribution to the equations
of motion is comparable to the fluxes everywhere in the compact
space.

We assume that the entire $H$-flux has one leg along the last
coordinate which we call the 9-direction i.e. $H=H_{ij9} dx^i \w
dx^j \w dx^9$. Furthermore, we assume that our space has a U(1)
isometry (at least in the smeared case) corresponding to shifts of
$x^9$. Then we can perform a T-duality
\cite{Buscher:1987sk,Hassan:1999mm} along this direction and find a
new space that can be conveniently written \cite{Shelton:2005cf} in
terms of the 1-forms $dx^\mu, dx^i, e^9 = dx^9 + \frac12 f^9_{ij} \,
x^i\, dx^j$ with $\mu = 0, \ldots,p-1$, $i=p, \ldots, 8$ and
$f^9_{ij} = H_{ij9}$. The T-dual setup has vanishing $H$-flux but
non-vanishing $f^9_{ij}$ which is often referred to as metric flux.
Note that $e^9$ is not closed but we rather have $\d e^9 = \frac12
f^9_{ij} dx^i \w dx^j$. Motivated by this T-duality, we consider an
O$p$-plane along $\mu=0,1,\ldots, p-1$ and the $e^9$ direction and
make the following Ansatz for the warped metric
\ba\label{eq:metricf}
\d s^2 &=& \d s^2_p + g_{99} e^9 e^9 + \d s^2_{9-p} = g_{\mu\nu} \d x^\mu \d x^\nu + g_{99} e^9 e^9 + g_{ij} \d x^i \d x^j \\
&=& \e^{2A} \lp \tilde{g}_{\mu\nu} \d x^\mu \d x^\nu +\tilde{g}_{99}
e^9 e^9 \rp + \e^{\frac{2(p+1)}{p-7}A} \tilde{g}_{ij} \d x^i \d
x^j,\nn
\ea
which is not block diagonal in the $dx$-basis since the
$e^9 e^9$ term gives contributions that mix the $\d x^9$ and $\d
x^i$ directions\footnote{Note that, unlike in a usual vielbein basis, $e^9$ does not have unit norm as $g_{99}$ is in general not equal to one. Furthermore, all tensors with an index 9 are always meant to be with respect to the basis form $e^9$ rather than $\d x^9$.}. Therefore, the Ricci tensor given in
\eqref{eq:ricci} is modified. It now has three contributions. One
from the unwarped metric $\tilde{g}_{ab}$, another from the metric
fluxes $f^9_{ij}$ and a third from the warp factor $A$. A lengthy
calculation leads to
\ba
R_{\mu\nu}&=& \tilde{R}_{\mu\nu} - \e^{\tfrac{16}{7-p}A} \tilde{g}_{\mu\nu} \, \tilde{\nabla}^2A \,,\nn \\
R_{99}&=& \tfrac12 \e^{\tfrac{32}{7-p}A} \tilde{g}_{99} \tilde{g}_{99} |\tilde{\d e^9}|^2 - \e^{\tfrac{16}{7-p}A} \tilde{g}_{99} \, \tilde{\nabla}^2A \,, \label{eq:riccif}\\
R_{9i} &=& \tilde{R}_{9i} + \tfrac{8}{7-p} \e^{\tfrac{16}{7-p}A} \tilde{g}_{99} |e^9 \w \d A \cdot \d e^9|_{9i}^\text{unwarped} \nn \\
R_{ij} &=& \tilde{R}_{ij} -\tfrac12 \e^{\tfrac{16}{7-p}A}
\tilde{g}_{99} |\tilde{\d e^9}|^2_{ij} - \tfrac{p+1}{p-7}
\tilde{g}_{ij} \tilde{\nabla}^2 A + \tfrac{8(p+1)}{p-7} \partial_i
A\partial_j A  \,,\nn
\ea
where $|A_n \cdot B_n|_{ab}^\text{unwarped} = \tfrac{1}{(n-1)!} A_{a c_1
\ldots c_{n-1}} B_{b d_1 \ldots d_{n-1}} \tilde{g}^{c_1 d_1}
\ldots \tilde{g}^{c_{n-1} d_{n-1}}$, and we assume that the metric \eqref{eq:metricf} has $\tilde{R}_{99}=0$. With this information we can now solve the equations of motions.

\subsection{The smeared solutions}
In this section we solve the equations of motions for a smeared O$p$-plane
i.e. in the non-warped case $A=0$ with $\delta(Op) \rightarrow 1$.
The non-zero fields are
\be\label{eq:F(8-p)}
\phi_0 = cst.,\quad F_{8-p} = m_{7-p} \w e^9,\quad R_{\mu\nu}, \quad R_{99}, \quad R_{9i}, \quad
R_{ij},
\ee
where $m_{7-p}$ is a closed $(7-p)$-form. The rest of
the RR-fields and the $H$-flux are zero. Again $p=1,\ldots,6$ and
some equations require minor modifications for $p=3$ due to the
self-duality of $F_5 = (1+\star)m_4 \w e^9$.

We start with the dilaton equation of motion for the smeared case,
\be\label{eq:source}
0 = \nabla^2 \phi_0 = \tfrac{p-3}{4}
\e^{\tfrac{p-3}{2}\phi_0}|F_{8-p}|^2- \tfrac{p-3}{4}
\e^{\tfrac{p-3}{4}\phi_0}\mu_p.
\ee
Using this in the external Einstein equation,
\be\label{eq:einsteinsmear}
R_{\mu\nu} = -\tfrac{7-p}{16}
\e^{\tfrac{p-3}{2}\phi_0} g_{\mu\nu} |F_{8-p}|^2 + \tfrac{7-p}{16}
\e^{\tfrac{p-3}{4}\phi_0} g_{\mu\nu}\mu_p\,,
\ee
one finds $R_{\mu\nu}=0$\footnote{\label{footnote}For the special case of $p=3$ we assume a
Ricci-flat external space, because this is no longer implied by \eqref{eq:source} and \eqref{eq:einsteinsmear}.}. This means that our setup only allows for $p$-dimensional Minkowski solutions.

The Bianchi identity for $F_{8-p}$ is
\be\label{eq:Bianchi}
\d F_{8-p} = -(-1)^p m_{7-p} \w \d e^9 = - \mu_p \epsilon_{9-p}.
\ee
Together with \eqref{eq:source} this gives
\be\label{eq:Fsquared}
-\star_{9-p} \d F_{8-p} = \e^{\tfrac{p-3}{4}\phi_0} |F_{8-p}|^2.
\ee
The Einstein equation in the 9-direction is
\ba\label{eq:R99}
R_{99} &=& \tfrac12 g_{99} g_{99} |\d e^9|^2 \nn \\
&=& \e^{\tfrac{p-3}{2}\phi_0}\lp \tfrac{1}{2} |F_{8-p}|_{99}^2 -\tfrac{7-p}{16} g_{99} |F_{8-p}|^2 \rp+ \tfrac{7-p}{16} \e^{\tfrac{p-3}{4}\phi_0} g_{99}\mu_p \\
&=& \e^{\tfrac{p-3}{2}\phi_0} \tfrac{1}{2} |F_{8-p}|_{99}^2,\nn
\ea
where we used \eqref{eq:source}. Since for $F_{8-p}=m_{7-p} \w e^9$
one has $|F_{8-p}|_{99}^2 =|F_{8-p}|^2 g_{99}$ we find
\be\label{eq:de9squared}
g_{99} |\d e^9|^2 = \e^{\tfrac{p-3}{2}\phi_0} |F_{8-p}|^2.
\ee
Together with $\eqref{eq:Fsquared}$ this leads to
\be
g_{99} \star_{9-p} \d e^9 \w \d e^9 = \e^{\tfrac{p-3}{2}\phi_0} g^{99}
\star_{9-p} m_{7-p} \w m_{7-p} = \e^{\tfrac{p-3}{4}\phi_0}
(-1)^p m_{7-p} \w \d e^9,
\ee
which implies
\be \label{eq:Ri9}
\boxed{\d e^9 = (-1)^p g^{99} \e^{\tfrac{p-3}{4} \phi_0}\star_{9-p}
m_{7-p}.}
\ee
This is of course nothing but the T-dual version of the
BPS condition \eqref{eq:ISDsmear}.

The Einstein equations for the directions transverse to the O$p$-plane are
\ba\label{eq:Rij}
R_{ij} &=& \tilde{R}_{ij} - \tfrac12 g_{99} |\d e^9|^2_{ij}\nn\\
&=&\e^{\tfrac{p-3}{2}\phi_0}\lp \tfrac{1}{2} |F_{8-p}|_{ij}^2 - \tfrac{7-p}{16}  g_{ij} |F_{8-p}|^2 \rp - \tfrac{p+1}{16} \e^{\tfrac{p-3}{4}\phi_0} g_{ij} \mu_p \\
&=&\e^{\tfrac{p-3}{2}\phi_0} \lp \tfrac{1}{2} |F_{8-p}|_{ij}^2- \tfrac{1}{2} g_{ij} |F_{8-p}|^2 \rp .\nn
\ea
Using \eqref{eq:Ri9} we can rewrite this as
\be
\tilde{R}_{ij} =\tfrac12 \e^{\tfrac{p-3}{2}\phi_0} g^{99} \lp |\star_{9-p} m_{7-p}|^2_{ij} + |m_{7-p}|_{ij}^2- g_{ij} |m_{7-p}|^2 \rp =0.
\ee
Note that $\tilde{R}_{ij}=0$ does not mean that we have no curvature since $R_{ij} \neq 0$ and $R_{99} \neq 0$. The final Einstein equation reads $\tilde{R}_{9i}=0$.

All other equations of motion are trivially satisfied, so that we have spelled out all the non-trivial equations of motion.

\subsection*{Summary of the solution}\label{summarysmeared}
The {\it a priori} non-zero fields in the Ansatz for a smeared O$p$-plane with $p=1,\ldots, 6$ and metric \eqref{eq:metricf} (with $A=0$) are
\begin{equation}
\phi_0 = cst.,\quad F_{8-p} = m_{7-p} \w e^9,\quad R_{\mu\nu}, \quad R_{99}, \quad R_{9i}, \quad R_{ij}.
\end{equation}
The equations of motion only allow for $p$-dimensional Minkowski
solutions ($R_{\mu\nu}=0$) modulo the caveat for $p=3$ mentioned in footnote \ref{footnote}. All Bianchi identities and equations of
motion can be reduced to $\tilde{R}_{9i}=0$ and the following two conditions:
\begin{itemize}
\item A duality condition between the curvature, which is encoded in the non-closure of the 1-form $e^9$, and the RR-flux
\begin{equation}\label{eq:duality}
\d e^9 = (-1)^p g^{99} \e^{\tfrac{p-3}{4} \phi_0}\star_{9-p} m_{7-p}\,.
\end{equation}
\item The amount of flux, and, as implied by \eqref{eq:duality}, also the curvature are fixed by the O$p$-plane charge,
\begin{equation}\label{eq:charge}
\mu_p = \e^{\tfrac{p-3}{4}\phi_0} |F_{8-p}|^2 =\e^{-\tfrac{p-3}{4}\phi_0}g_{99} |\d e^9|^2\,.
\end{equation}
\end{itemize}
After giving a simple example we will proceed to show that these
solutions can be localised by introducing a warp factor and allowing
the dilaton to vary.

\subsection{A simple example}
There are many examples of so called twisted tori and coset spaces
that can be used to obtain explicit Minkowski solutions of the type
described above. Probably the simplest case is the one where the
metric fluxes satisfy the Heisenberg algebra. Such a space can be
easily compactified and is then T-dual to a three-torus with $H$-flux
\cite{Douglas:2006es}. To obtain a simple solution we
choose the unwarped metric to be $\tilde{g}_{\mu\nu}=\eta_{\mu\nu}$, $\tilde{g}_{99}=1$,
and $\tilde{g}_{ij} = \delta_{ij}$. An explicit example that solves the equations of
motion is then an O$p$-plane with $p=1,\ldots,6$ that wraps the
directions $0,1, \ldots, p-1,9$, and the directions $p, p+1, \ldots,
8$ and 9 are compact. Furthermore, we choose the non-zero fields for
our solution to be
\ba
\phi_0&=&cst.,\,\, F_{8-p} = (-1)^p \e^{-\tfrac{p-3}{4}\phi_0} f\, \d x^p \w \d x^{p+1} \w \ldots \w \d x^6 \w e^9,\\
f^9_{78}&=&-f^9_{87} = f \,\, \Rightarrow \,\,R_{77}=R_{88} =
-R_{99} = -\tfrac12 f^2, \quad \text{with} \quad f^2 =
\e^{\tfrac{p-3}{4}\phi_0} \mu_p.\nn
\ea
This is a $p$-dimensional
Minkowski solution in which a smeared O$p$-plane is compactified on
an internal everywhere negatively curved space. As we explain in
the following subsection, the O$p$-plane can be localised if we
introduce a warp factor and allow the dilaton to vary over the
internal space.

\subsection{The localised solutions }
To find localised solutions with $p =1, \ldots, 6$ we now include a
non-zero warp-factor $A$ which we allow together with the dilaton
$\phi$ to depend on all the coordinates transverse to the O$p$-plane. For $F_{8-p}$  we make the Ansatz
\be
F_{8-p} = \hat{F}_{8-p}
- \e^{-2(p+1)A-\tfrac{p-3}{2} \phi} \tilde{\star}_{9-p} \d \alpha,
\ee
where $\hat{F}_{8-p} =m_{7-p} \w e^9$ is the $(8-p)$-form from \eqref{eq:F(8-p)} and $\alpha$ is a function of the transverse coordinates that will be determined below. The tilde will always mean the unwarped metric $\tilde{g}_{ab}$, the corresponding unwarped Hodge star $\tilde{\star}$, or contractions of forms done with $\tilde{g}_{ab}$.\\
We start out by deriving a BPS condition similar to \eqref{BPS}. The dilaton equation of motion
\ba\label{eq:sourcelocal}
\nabla^2 \phi &=&\e^{\tfrac{2(p+1)}{7-p}A} \tilde{\nabla}^2 \phi \\
&=& \tfrac{p-3}{4} \e^{\tfrac{p-3}{2}\phi}\lp |\hat{F}_{8-p}|^2 +
\e^{\tfrac{2(p+1)(p-6)}{7-p}A-(p-3)\phi}(\tilde{\partial}\alpha)^2
\rp - \tfrac{p-3}{4} \e^{\tfrac{p-3}{4}\phi}\mu_p \delta(Op), \nn
\ea
gives us an expression for $\tilde{\nabla}^2 \phi$, and we
find $\tilde{\nabla}^2 A$ from the trace of the Einstein equations
along the O$p$-plane,
\ba\label{eq:traceEinstein}
\tilde{g}^{\mu\nu} R_{\mu\nu} + \tilde{g}^{99} R_{99} &=& \tilde{R}_{p} - \e^{\tfrac{16}{7-p}A} (p+1) \tilde{\nabla}^2 A + \tfrac12 \e^{\tfrac{32}{7-p}A} \tilde{g}_{99} |\tilde{\d e^9}|^2 \nn \\
&=& -\tfrac{(7-p)(p+1)}{16} \e^{2A+\tfrac{p-3}{2}\phi} \lp |\hat{F}_{8-p}|^2+ \e^{\tfrac{2(p+1)(p-6)}{7-p}A-(p-3)\phi}(\tilde{\partial}\alpha)^2\rp \\
&+& \tfrac{1}{2} \e^{\tfrac{p-3}{2}\phi}|\hat{F}_{8-p}|_{99}^2 +
\tfrac{(7-p)(p+1)}{16} \e^{2A+\tfrac{p-3}{4}\phi} \mu_p \delta(Op)\,.\nn
\ea
Finally, from the Bianchi identity for $F_{8-p}$,
\ba\label{eq:Bianchi2}
\d F_{8-p} &=& \d \hat{F}_{8-p} - \d \lp \e^{-2(p+1)A-\tfrac{p-3}{2} \phi} \tilde{\star}_{9-p} \d \alpha \rp = -(-1)^p m_{7-p} \w \d e^9 \\
&& - (-1)^{p} \lp \e^{\tfrac{(p-5)(p+1)}{7-p}A-\tfrac{p-3}{2}\phi} \tilde{\nabla}^2 \alpha + \e^{\tfrac{(9-p)(p+1)}{7-p}A} \tilde{g}^{ij} \partial_i \lp \e^{-2(p+1)A-\tfrac{p-3}{2}\phi}\rp \partial_j \alpha\rp \epsilon_{9-p} \nn \\
&=& -\mu_p \delta(Op) \epsilon_{9-p}\,, \nn
\ea
we get $\tilde{\nabla}^2 \alpha$. Putting all these expressions together, we have
\ba\label{eq:BPStwistedtori}
&&\tilde{\nabla}^{2}\Big(\e^{(p+1)A+\frac{p-3}{4}\phi} + (-1)^{p} \alpha \Big) =  \e^{\frac{\left(p-3\right)^{2}}{p-7}A+\tfrac{p-3}{4} \phi}\tilde{R}_{p} \nn\\
&+& \e^{\tfrac{(p+1)(9-p)}{p-7}A-\tfrac{p-3}{4}\phi}\left\vert \partial\left( \e^{(p+1)A+\tfrac{p-3}{4}\phi}+ \left(-1\right)^{p}\alpha \right)  \right\vert^{2} \\
&+& \tfrac{1}{2} \e^{\tfrac{(p+1)(p-5)}{p-7}
A+\tfrac{3(p-3)}{4}\phi} \left\vert \sqrt{g^{99}} m_{7-p}- (-1)^p
\e^{-\tfrac{p-3}{4}\phi} \sqrt{g_{99}} \star_{9-p} \d
e^9\,\right\vert^{2}\,,\nn
\ea
where the squares in the last two term are with respect to the warped metric.
Integrating both sides over the compact space we find for Minkowski solutions
that
\be\label{eq:alpha}
\alpha = (-1)^{p+1} \e^{(p+1)A+\tfrac{p-3}{4}\phi}+ cst.
\ee
and
\be
\d e^9 = (-1)^p \e^{\tfrac{p-3}{4}\phi} g^{99} \star_{9-p}
m_{7-p}.
\ee
Plugging the dilaton equation into
the external Einstein equation we furthermore find
\be
0 = \tfrac{4(p-3)}{7-p} \tilde{R}_{\mu\nu} =
\tilde{g}_{\mu\nu} \e^{\tfrac{16}{7-p}A} \tilde{\nabla}^2 \lp
\tfrac{4(p-3)}{7-p}A - \phi \rp\,.
\ee
Since a harmonic function on a compact space is constant, we have
\be\label{eq:dilaton}
\phi= \tfrac{4(p-3)}{7-p} A +\phi_0,
\ee
which implies that the duality condition \eqref{eq:Ri9} is unchanged since
\be\label{eq:dualitylocal}
\d e^9 = (-1)^p \e^{\tfrac{p-3}{4}\phi} g^{99} \star_{9-p}
m_{7-p} = (-1)^p \e^{\tfrac{p-3}{4}\phi_0} \tilde{g}^{99}
\tilde{\star}_{9-p} m_{7-p}.
\ee

Next we check the internal Einstein equations
\ba
R_{ij} &=& \tilde{R}_{ij} -\tfrac12 \e^{\tfrac{16}{7-p}A} \tilde{g}_{99} |\tilde{\d e^9}|^2_{ij} - \tfrac{p+1}{p-7} \tilde{g}_{ij} \tilde{\nabla}^2 A + \tfrac{8(p+1)}{p-7} \partial_i A \partial_j A \nn\\
&=& \tfrac12 \partial_i \phi \partial_j \phi +\e^{\tfrac{p-3}{2}\phi} \Bigg[ \tfrac{1}{2} |\hat{F}_{8-p}|_{ij}^2 + \tfrac{1}{2} \e^{-2(p+1) A-(p-3) \phi}\lp \tilde{g}_{ij} (\tilde{\partial} \alpha)^2-\partial_i \alpha \partial_j \alpha \rp \nn\\
&& - \tfrac{7-p}{16} \e^{\tfrac{2(p+1)}{p-7}A} \tilde{g}_{ij} \lp
|\hat{F}_{8-p}|^2 +
\e^{\tfrac{2(p+1)(p-6)}{7-p}A-(p-3)\phi}(\tilde{\partial}\alpha)^2 \rp \Bigg] \\
&&- \tfrac{(p+1)}{16} \e^{\tfrac{2(p+1)}{p-7}A+\tfrac{p-3}{4}\phi}
\tilde{g}_{ij} \mu_p \delta(Op),\nn
\ea
which using \eqref{eq:charge}, \eqref{eq:traceEinstein} and \eqref{eq:dilaton}
reduces to \eqref{eq:Rij}.

There is one more non-trivial Einstein equation:
\ba
R_{9i} &=& \tfrac{8}{7-p} \e^{\tfrac{16}{7-p}A} \tilde{g}_{99} |e^9 \w \d A \cdot \d e^9|_{9i}^\text{unwarped} \\
&=&\tfrac{8}{7-p} \e^{\tfrac{16}{7-p}A+\tfrac{p-3}{4}\phi_0} (-1)^p
|e^9 \w \d A \cdot \tilde{\star}_{9-p} m_{7-p}|_{9i}^\text{unwarped}
\,,\nn
\ea
which is satisfied due to \eqref{eq:dualitylocal}.\\
Finally, we have to make sure that the equation of motion for $F_{8-p}$ is satisfied i.e.
\ba\label{eq:eomF}
0&=&\d \lp \e^{\tfrac{p-3}{2} \phi} \star_{10} F_{8-p} \rp = \d \lp \e^{\tfrac{p-3}{2}\phi} \star_{10} \hat{F}_{8-p}\rp - \d \lp (-1)^p \tilde{\star}_{p+1} 1 \w \d \alpha \rp \\
&=&\tfrac{16}{7-p} \e^{\tfrac{16}{7-p} A+\tfrac{p-3}{2}
\phi_0} \d A \w \epsilon_p \w \lp -(-1)^p \sqrt{\tilde{g}^{99}}
\tilde{\star}_{9-p} m_{7-p} + \sqrt{\tilde{g}_{99}}
\e^{-\tfrac{p-3}{4} \phi_0} \d e^9 \rp\,,\nn
\ea
where we have used \eqref{eq:alpha} and \eqref{eq:dilaton}. This
equation is again satisfied due to \eqref{eq:dualitylocal}.

All other equations of motion are trivially satisfied, so that we
have spelled out all the non-trivial equations of motions.

As an important observation, we note, using \eqref{eq:metricf} and \eqref{eq:riccif}, that
\be
\int \sqrt{g^{(10)}}R_{(10-p)}= \int \sqrt{\tilde g} \left\{{ - 8\frac{p+1}{7-p} \left({}\right.\!\! \tilde \nabla A\!\!\left.{}\right)^2 - \frac{1}{4} e^{\frac{16}{7-p}A} \tilde{g}_{99} \tilde g^{ij} \tilde g^{kl} f^9_{ik}f^9_{jl} }\right\} <0,
\ee
where the integral is over the $(10-p)$-dimensional internal space, and we have dropped total derivative terms. This is the integral of the full internal curvature computed from the warped metric components $g_{99}$ and $g_{ij}$ weighted with the full \emph{ten}-dimensional metric determinant. This quantity is, up to a Weyl-rescaling, just minus the contribution of the internal curvature to the $p$-dimensional scalar potential \footnote{The inclusion of the $p$-dimensional part of the metric determinant is, in fact, important for the manifestly negative sign, as it cancels some remaining warp factors in front of the total derivative terms.}. The last term describes the
original negative curvature term of the twisted torus, now dressed with a warp factor. The first term is
due to the non-constancy of the warp factor in the localised case and clearly vanishes in the smeared limit. Interestingly, both terms are manifestly negative (assuming $p<7$, as we do in this paper), even though the negative tension objects are now properly localised.
This somewhat circumvents difficulties found in \cite{Douglas:2010rt}, where negative curvature spaces for uplifting potentials were argued to be problematic as the negative energy-momentum sources supporting them are really localized objects. It should be emphasized, however, that there is no real uplifting in the present context, as we are looking at Minkowski BPS solutions.

\subsection*{Summary of the solution}
We have shown that the smeared solution of subsection
\ref{summarysmeared} persists after localisation, if we introduce a
warp factor $A$ and add a new term to the RR field strength
$F_{8-p}$,
\be
F_{8-p} = \hat{F}_{8-p} -
\e^{-2(p+1)A-\tfrac{p-3}{2}\phi} \tilde{\star}_{9-p} \d \alpha =
m_{7-p} \w e^9 +\tfrac{16}{7-p} (-1)^p \e^{-\tfrac{16}{7-p}
A-\tfrac{p-3}{4} \phi_0} \tilde{\star}_{9-p} \d A\,.
\ee
The only modulus that has changed is the dilaton that is no longer constant
but rather proportional to the warp-factor
\be
\phi=\frac{4(p-3)}{7-p} A + \phi_0.
\ee
The warp factor is
determined for example through \eqref{eq:sourcelocal} which becomes for our
choice of $\alpha$ \eqref{eq:alpha}
\be\label{eq:warping}
\tilde{\nabla}^2 \e^{\tfrac{16}{p-7} A} = -\e^{\tfrac{p-3}{2} \phi_0}
|\tilde{\hat{F}}_{8-p}|^2 + \e^{\tfrac{p-3}{4} \phi_0} \mu_p
\tilde{\delta}(Op),
\ee
where we have pulled out the warp factor
dependence in $|\tilde{\hat{F}}_{8-p}|^2$ and $\tilde{\delta}(Op)$.

In \cite{Douglas:2010rt} the authors have shown that
compactifications down to dS or Minkowski space on everywhere
negatively curved spaces must have significant warping and/or large
stringy corrections. Since we have neglected stringy corrections but
found compactifications that lead to Minkowski solutions, we can
conclude that we have large warping everywhere. This means that the
contributions of the warp factor are of the same order as the
contributions of the fluxes and localised sources. This is apparent
from \eqref{eq:warping}. Far away from a localised source one could
have naively expected that the warp factor approaches a constant and
the warping becomes negligible. This would have made it impossible
to find compactifications on negative curvature spaces that have a
large volume i.e. in a regime where we can trust supergravity.
However, as we can see from \eqref{eq:warping} the warp factor is
everywhere sourced by the fluxes so that its contributions will be
relevant everywhere even if the compact space is large.

As we saw in our particular example, however, the warping and the anisotropic conformal rescalings of the internal metric can conspire to still give a curvature contribution to the $p$-dimensional scalar potential that is manifestly positive, showing that the situation can be more complex than expected from \cite{Douglas:2010rt}.

\section{Non-BPS solutions}\label{nonBPS}
In the previous two sections we saw that the localization of smeared sources that satisfy a
 BPS condition  leads to modifications of the solution, which, although important individually,
 still leave many features of the smeared solution unchanged when they are added up.
 As we will now show, in a non-BPS situation localisation corrections may in general lead to much more drastic effects.

\subsection{The smeared solutions}
To begin with, we look for non-BPS solutions with smeared sources,
assuming that the dilaton is constant and the metric has the form
of a direct product,
\begin{equation}
\phi=\phi_0\,,\qquad \d s^2_{10}=\d s_{p+1}^2 + \d s_{9-p}^2\,.
\end{equation}
The non-zero form fields are taken to be $H$ and $F_{6-p}$ with $p=2,\ldots,6$, with the rest of the
RR-fields being identically zero. It is convenient to pull out the
part of $F_{6-p}$ that is along $\star_{9-p} H$:
\be
F_{6-p} = \bar{F}_{6-p} + (-1)^p \e^{-\tfrac{p+1}{4}\phi_0} \kappa \star_{9-p}
H,\,
\ee
where $\bar{F}_{6-p}$ is a closed and co-closed form, satisfying $\bar{F}_{6-p}\w H = 0$.

The Bianchi identity for $F_{8-p}=0$ then becomes
\begin{equation}
0= \e^{-\tfrac{p+1}{4}\phi_0} \kappa |H|^2 \,\ep_{9-p} -
\mu_p\,\ep_{9-p}\,.
\end{equation}
So we have $\e^{-\tfrac{p+1}{4}\phi_0} \kappa |H|^2 = \mu_p$ and
therefore $\kappa >0$ for an orientifold plane and $\kappa<0$ for a
D-brane corresponding to $\mu_p<0$. Using the $F_{8-p}$ Bianchi identity in the dilaton equation of motion one finds
\be \label{eq:F6p}
|\bar{F}_{6-p}|^2 = \lp -\kappa^2
+\tfrac{p-3}{p-1} \k +\tfrac{2}{p-1}\rp \e^{-\tfrac{p+1}{2} \phi_0}
|H|^2.
\ee
Since $|\bar{F}_{6-p}|^2$ and $|H|^2$ are internal and
therefore positive we find the range $-\tfrac{2}{p-1} \leq \k \leq
1$. The external Einstein equation simplifies to
\be\label{eq:EinsteinEx}
R_{\mu\nu} = -\tfrac{1-\k}{2(p-1)} \e^{-\phi_0} |H|^2 g_{\mu\nu}.
\ee
So we see that we have AdS solutions for $\k<1$ (the case $\kappa =1$ corresponds to the BPS Minkowski solutions discussed in section \ref{MinkowskiRicciflat}). The internal Einstein
equation gives
\be
R_{ij} = -\tfrac{(1-\k)(1+\k(p-1))}{2(p-1)} e^{-\phi_0} |H|^2 g_{ij} +
\tfrac12 \e^{\frac{p-1}{2} \phi_0}
|\bar{F}_{6-p}|^2_{ij}+\tfrac12(1-\k^2) \e^{- \phi_0} |H|^2_{ij},
\ee
assuming $|\bar{F}_{6-p} \cdot H|_{ij} = 0$. Although the internal components of the Ricci tensor have no
fixed sign, one finds from taking the trace
\be
R_{9-p} = \tfrac{(1-\k)p}{p-1} e^{-\phi_0} |H|^2 = -\tfrac{2p}{p+1} R_{p+1}.
\ee
Since the external space is AdS the internal space has to be positively curved.

\subsection{A simple example}
Let us consider the simple situation $\kappa=-\tfrac{2}{p-1}$, for which \eqref{eq:F6p}
implies $\bar{F}_{6-p}=0$, and we furthermore have a net D-brane charge,
\be\label{mu}
\mu_p=-\tfrac{2}{p-1}\e^{-\tfrac{p+1}{4}\phi_0}|H|^2\,.
\ee
The value of the external Ricci scalar becomes
\be \label{cc}
R_{p+1}=-\tfrac{(p+1)^2}{2(p-1)^2}\e^{-\phi_0}|H|^2\,,
\ee
and the internal Einstein equation reduces to
\be
R_{ij}=\tfrac{p+1}{2(p-1)^2}\e^{-\phi_0}|H|^2 g_{ij} + \tfrac{(p+1)(p-3)}{2(p-1)^2}\e^{-\phi_0}|H|^2_{ij}\,.
\ee
To find simple explicit solutions we take the internal space to be a product of two spheres (cf. \cite{Silverstein:2004id} for $p=3$),
\be
M_{9-p}=S^3\times S^{6-p}\,,
\ee
with volume forms $\epsilon_3$ and $\epsilon_{6-p}$. Then the solution
reads
\ba
H &=& h \epsilon_3\,,\\
F_{6-p} &=& -\tfrac{2}{p-1}(-1)^p \e^{-\tfrac{p+1}{4}\phi_0} \star_{9-p} H = -\tfrac{2}{p-1} \e^{-\tfrac{p+1}{4}\phi_0} h \epsilon_{6-p}\,,
\ea
and the internal Einstein equation fixes the curvature radii (and hence
the volume) of the $S^3$ and the $S^{6-p}$,
\ba
R^{S^3}_{ij}&=& \tfrac{(p+1)(p-2)}{2(p-1)^2}\e^{-\phi_0} h^2 g^{S^3}_{ij}\,, \\
R^{S^{6-p}}_{ij} &=& \tfrac{p+1}{2(p-1)^2}\e^{-\phi_0} h^2 g^{S^{6-p}}_{ij}\,.
\ea

\subsection{The failure to localise?}

In order to argue that non-BPS solutions based on smeared sources should in general not be localisable, we study an explicit example.
We show this for $\kappa = -\tfrac{2}{p-1}$, since then $\bar{F}_{6-p} =0$ due to \eqref{eq:F6p}\footnote{For an interesting discussion of local solutions that are close to but not at the BPS point see \cite{Baumann:2010sx}.}.

We will first consider the case with $p=3$ since we can easily compare the O3-plane
solution of GKP \cite{Giddings:2001yu} with $\kappa=1$, i.e. ISD flux, to a solution with
a D3-brane and $\kappa=-1$, i.e. IASD fluxes.  To this end, we write for the $F_3$-flux
\begin{equation}\label{duality}
F_3 = \mp \e^{-\phi} \star_6 H\,,
\end{equation}
where from here on the upper sign corresponds to the BPS solution and the lower sign to the non-BPS solution. Equation \eqref{duality} is our starting point as we take it as the definition of the solution. Whether a less standard Ansatz might lead to localised solutions that reduce, in the smeared limit, to the smeared solutions constructed above is not clear. One should however keep in mind that \eqref{duality} fixes some of the moduli at certain values, so changing this condition could change the entire solution in a more non-trivial way. We leave this for future investigation \cite{progress2}.

As in the smeared solution, the fluxes satisfy the Bianchi identities
\begin{equation}
\d H=\d F_3=0\,.
\end{equation}
The most general metric Ansatz for the localised solution is
\begin{equation}
\d s_{10}^2 =\e^{2A}\d \tilde{s}_4^2 + \d s^2_6\,,
\end{equation}
where the internal metric $\d s_6^2$ is left arbitrary.
With \eqref{duality}, the integrated dilaton equation of motion implies that $F_1$ vanishes, as there are no net orientifold charges for $p\neq 3$ in our setup. The dilaton equation then also implies that $\phi=\phi_{0}$ is constant.
The Ansatz for $F_5$ reads\footnote{One can verify that this Ansatz is required by the self-duality of $F_5$ and the equations of motion for $F_3$ and $H$, which also imply that there cannot be any warp factor appearing in \eqref{duality}.}
\begin{equation}
F_5=-(1+\star_{10})\e^{-4A} \star_6 \d\alpha\,.
\end{equation}

Let us go through the equations of motion. The dilaton equation is
automatically satisfied due to the duality condition
(\ref{duality}). The equations of motion for $H$ and $F_3$ are
solved if
\begin{equation}
\alpha=\pm \e^{4A} +cst.
\end{equation}
The different sign for the $F_5$ field in both cases makes sense
since the source is an O3-plane in the BPS case and a D3-brane in
the non-BPS case. The $F_5$ equation then gives
\begin{equation}
4 \nabla^2  A  = \e^{-\phi_0} |H|^2 \mp \mu_3 \delta(O3/D3)\,.
\end{equation}
Note, that $\mu_3>0$ for the O3-plane and $\mu_3<0$ for the D3-brane. The external Einstein equation is
\begin{equation}
\e^{-2A}\tilde{R}_4- 4\nabla^2 A = -\e^{-\phi_0} |H|^2 +\mu_3 \delta(O3/D3)\,.
\end{equation}
Combining these two equations leads to
\begin{equation}\label{important}
\boxed{\e^{-2A} \tilde{R}_4 = (1 \mp 1)\mu_3 \delta(O3/D3)}
\end{equation}
Clearly, in the BPS case, we recover the Minkowski solution
$\tilde{R}_4=0$ while in the non-BPS case there is no AdS solution
anymore! One simply goes away from the source and then this equation
is inconsistent with an AdS solution ($\tilde{R}_4=-2|\Lambda|$).
Even if one tries to make sense out of
this by regularising the delta function, one needs that
$\e^{-2A}\sim \delta_{\text{regularised}}(D3)$, which is not true.
Only in the smeared limit, where
\begin{equation}
\delta(D3) \rightarrow 1\,,\qquad A\rightarrow 0\,,
\end{equation}
equation (\ref{important}) makes sense since we reproduce the
previous result (\ref{mu}), (\ref{cc}). Due to our mild assumptions
we believe that our simple smeared solution ceases to exist
once localisation effects are taken into account. We should point out
that a completely analogous localisation problem is encountered for
the corresponding AdS solution with ISD fluxes and an anti-D3 brane.

For $p\neq 3$ one can likewise show that the smeared solutions with $\kappa = -\tfrac{2}{p-1}$, i.e. $\bar{F}_{6-p} =0$, have to get altered or even disappear upon taking the localisation effects into account: plugging the Ansatz $F_{6-p} = -\tfrac{2}{p-1} (-1)^p \e^{-\tfrac{p+1}{4}\phi} \star_{9-p} H$ into the $F_{6-p}$ and $H$ equations of motion for a generic $F_{8-p}$ and dilaton $\phi$ leads to a contradiction.

\begin{figure}[h!]
\center
\includegraphics[width=0.7\textwidth]{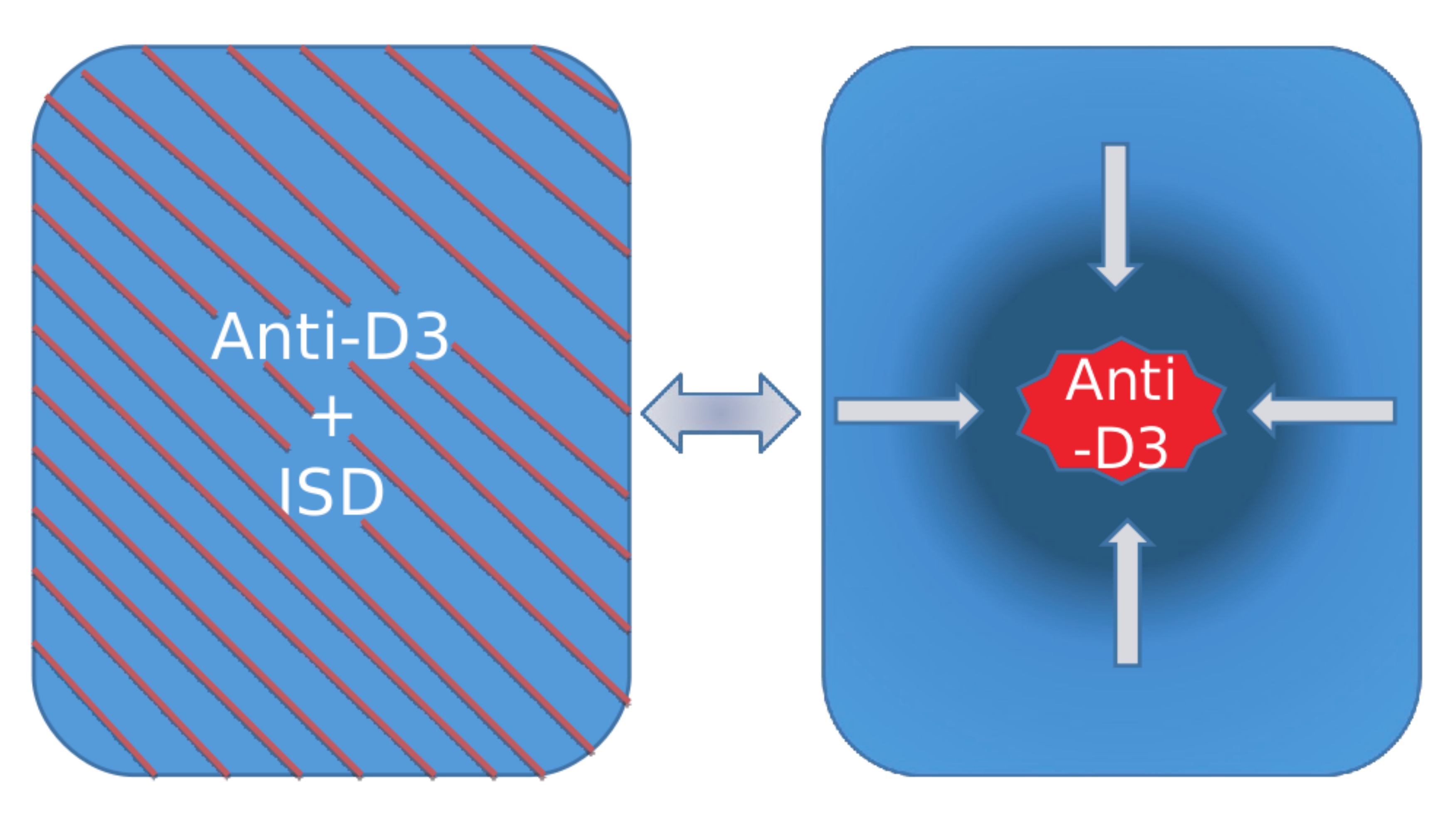}
\caption{The smeared (left) versus localised (right) case.\label{fig:fig1}}
\end{figure}

There is a simple physical picture behind these technical difficulties with localisation. As is well known I(A)SD fluxes act as \emph{smeared} (anti-) D3 branes, from the point of view of their charge and energy-momentum. Therefore adding anti-D3 branes to ISD flux (or vice versa), creates a perturbative instability when we localise the anti-branes, as shown in figure \ref{fig:fig1}. When the anti-branes are localised they single out a preferred point that attracts ISD fluxes. In the smeared case there is no preferred point of attraction, and the instability is only non-perturbative (brane-flux annihilation).

Even though our localised Ansatz is not the most general one (see discussion below (\ref{duality})), this physical reasoning seems to strengthen our belief in that even the most general Ansatz would fail to give a solution.

\section{Discussion}\label{Discussion}
We have presented BPS solutions from compactifications on Ricci-flat and negatively curved spaces.
The T-duality chain that relates these BPS solutions to each other is
quite straightforward if one T-dualises the \emph{smeared} GKP
solution on a torus \footnote{The Buscher rules
\cite{Buscher:1987sk, Hassan:1999mm} need a $U(1)$ symmetry, which
requires us to smear the solutions.}. To obtain the solutions with
a Ricci-flat internal space one either takes the T-duality circle along
the orientifold (going down in dimension) or on the torus along a
cycle without $H$-flux. If we T-dualise a cycle with $H$-flux, we obtain a
solution on a twisted torus. The solutions obtained in this way can
again be T-dualised up or down giving rise to all the solutions we
have presented. Characterising the solutions by their
$D$-dimensional generalisation of the ISD condition, we have schematically
\be
\boxed{H \propto \star_{9-p} F_{6-p}} \quad
\underset{T_{F_{7-p}}}{\overset{T_{Op}}{\rightleftarrows}} \quad \boxed{H
\propto \star_{10-p} F_{7-p}} \quad
\underset{T_{e^9}}{\overset{T_{H}}{\rightleftarrows}} \quad \boxed{\d e^9
\propto\star_{9-p} i_9 F_{8-p}}\,,
\ee
where the first pair of arrows
indicates how we connect the various smeared solutions with Ricci-flat
internal spaces and the second
arrow pair connects the solutions with Ricci-flat internal spaces to the
twisted tori. The latter solutions are characterised by a duality
condition between the metric flux and the RR-flux.

By going through the equations of motion we demonstrated that the smeared BPS
solutions can be localised\footnote{Note that this is more general
than arguments that rely on integrability from supersymmetry. As in
GKP \cite{Giddings:2001yu} we also allow the solutions to break
SUSY.} and that the moduli do not shift (even for the smeared solutions
with everywhere negative internal curvature). This is attributed to the
BPS condition that makes the contributions from the localisation
cancel against each other in the effective potential \cite{DeWolfe:2002nn}. An intuitive argument
for this is the ``no-force'' condition for mutually BPS objects. We have the ISD fluxes (and
their T-dual generalisations) which act as a smeared source with positive
charge and positive tension and the O$p$-plane. General lore says that when such objects are combined
in a mutually BPS way they will not affect each other due to the cancellation
between gravitational forces and electromagnetic forces.

While we believe that solutions that do not get altered will be BPS, we do not claim that all
BPS solutions with smeared orientifolds allow a localisation. It rather
seems that the localisation does not always work in such a simple way
for BPS solutions. It is for
example not known whether or how the supersymmetric solutions
of \cite{DeWolfe:2005uu} are localisable, mainly because they
involve multiple intersecting orientifolds. Localised solutions
with similarly BPS intersecting O$p$-planes or D$p$-branes are not
even known in ten-dimensional flat space.

We also studied simple non-BPS solutions where one generically
expects problems when trying to localise smeared sources. Indeed, our most
important result concerns the non-BPS AdS solutions, which we have
derived in the smeared limit\footnote{In $D=4$ these are
solutions with non-ISD fluxes.}. To our knowledge this is
the first time that such explicit solutions have been constructed.
We have argued from a 10D point of view that the localisation
procedure fails for these solutions so that they probably cease to exist.

If one studies flux compactifications from the point of view of an
effective potential, one has to estimate the size of the
\emph{individual} localisation corrections (such as the warping
correction). The separate terms (the one from the warp factor)
were argued in \cite{Douglas:2010rt} to be of the same order as
the fluxes when one compactifies on everywhere negatively curved
spaces. This was one of the motivations for the explicit presentation of the twisted tori solutions in this paper. We have generalised the existing localised orientifold solutions on twisted tori in two ways. First, we extended to Minkowski vacua of different dimensions. Second, our solutions are BPS-like but not necessarily SUSY, because we have analysed the full 10D equations of motion, instead of just the pure spinor equations. As an important application, we found that the properly integrated internal curvature stays negative,
even after localisation of the orientifold planes (in fact, the warp factor gradients even
introduce an additional negative term in the integrated internal
curvature).
We would like to point out that T-duality connects the twisted tori solutions to the GKP-like solutions. Therefore the arguments of \cite{Douglas:2010rt} extends to Ricci-flat internal spaces with $H$-flux, even in the large volume
limit. Regardless of the size of the warping, we have shown that the sum of the
localisation corrections cancels in certain cases at the BPS
points. However, we expect that smeared non-BPS solutions get
changed or might even cease to exist when the localisation effects
are taken into account. Concretely, what we have shown in an explicit example
is that there is an incompatibility between having a static
solution (stable or unstable) based on mutually non-BPS building
blocks solving all the equations of motion, and sensible
localisation. It would be interesting to understand in more
detail how strong the back reaction of the localised source is,
perhaps in the style of the investigations done in \cite{Bena:2009xk}.

\section*{Acknowledgements}
We like to thank Ralph Blumenhagen, Steve Giddings, Thomas Grimm,
Michael Haack, Shamit Kachru, Paul Koerber, Luca Martucci, Liam
McAllister, Gary Shiu and Jian Qiu for useful discussions and Bret
Underwood for discussions and valuable comments on an earlier
draft. U.D. is supported by the Swedish Research Council (VR) and
the G\"oran Gustafsson Foundation. D.J., T.W. and M.Z. are
supported by the German Research Foundation (DFG) within the Emmy
Noether Program (Grant number ZA 279/1-2) and the Cluster of
Excellence ``QUEST''. T.V.R. is supported by the G\"oran
Gustafsson Foundation. T.W. is supported by the Alfred P. Sloan
Foundation and by the NSF under grant PHY-0757868. This research
was supported in part by the National Science Foundation under
Grant No. NSF PHY05-51164.

\bibliography{groups}

\providecommand{\href}[2]{#2}\begingroup\raggedright\begin{thebibliography}{10}

\bibitem{Giddings:2001yu}
S.~B. Giddings, S.~Kachru and J.~Polchinski,  {\em {Hierarchies from fluxes in
  string compactifications}}, Phys. Rev. {\bf D66} (2002) 106006
[\href{http://www.arXiv.org/abs/hep-th/0105097}{{\tt hep-th/0105097}}].

\bibitem{Schulz:2004ub}
M.~B. Schulz,  {\em {Superstring orientifolds with torsion: O5 orientifolds of
  torus fibrations and their massless spectra}}, Fortsch. Phys. {\bf 52} (2004)
  963--1040
[\href{http://www.arXiv.org/abs/hep-th/0406001}{{\tt hep-th/0406001}}].

\bibitem{Grana:2006kf}
M.~Gra\~na, R.~Minasian, M.~Petrini and A.~Tomasiello,  {\em {A scan for new
  N=1 vacua on twisted tori}}, JHEP {\bf 05} (2007) 031
[\href{http://www.arXiv.org/abs/hep-th/0609124}{{\tt hep-th/0609124}}].

\bibitem{Angelantonj:2003rq}
C.~Angelantonj, S.~Ferrara and M.~Trigiante,  {\em {New D = 4 gauged
  supergravities from N = 4 orientifolds with fluxes}}, JHEP {\bf 10} (2003)
  015
[\href{http://www.arXiv.org/abs/hep-th/0306185}{{\tt hep-th/0306185}}].

\bibitem{Derendinger:2004jn}
J.-P. Derendinger, C.~Kounnas, P.~M. Petropoulos and F.~Zwirner,  {\em
  {Superpotentials in IIA compactifications with general fluxes}}, Nucl. Phys.
  {\bf B715} (2005) 211--233
[\href{http://www.arXiv.org/abs/hep-th/0411276}{{\tt hep-th/0411276}}].

\bibitem{Randall:1999ee}
L.~Randall and R.~Sundrum,  {\em {A large mass hierarchy from a small extra
  dimension}}, Phys. Rev. Lett. {\bf 83} (1999) 3370--3373
[\href{http://www.arXiv.org/abs/hep-ph/9905221}{{\tt hep-ph/9905221}}].

\bibitem{DeWolfe:2002nn}
O.~DeWolfe and S.~B. Giddings,  {\em {Scales and hierarchies in warped
  compactifications and brane worlds}}, Phys. Rev. {\bf D67} (2003) 066008
[\href{http://www.arXiv.org/abs/hep-th/0208123}{{\tt hep-th/0208123}}].

\bibitem{Giddings:2005ff}
S.~B. Giddings and A.~Maharana,  {\em {Dynamics of warped compactifications and
  the shape of the warped landscape}}, Phys. Rev. {\bf D73} (2006) 126003
[\href{http://www.arXiv.org/abs/hep-th/0507158}{{\tt hep-th/0507158}}].

\bibitem{Martucci:2006ij}
L.~Martucci,  {\em {D-branes on general N = 1 backgrounds: Superpotentials and
  D-terms}}, JHEP {\bf 06} (2006) 033
[\href{http://www.arXiv.org/abs/hep-th/0602129}{{\tt hep-th/0602129}}].

\bibitem{Frey:2006wv}
A.~R. Frey and A.~Maharana,  {\em {Warped spectroscopy: Localization of frozen
  bulk modes}}, JHEP {\bf 08} (2006) 021
[\href{http://www.arXiv.org/abs/hep-th/0603233}{{\tt hep-th/0603233}}].

\bibitem{Koerber:2007xk}
P.~Koerber and L.~Martucci,  {\em {From ten to four and back again: how to
  generalize the geometry}}, JHEP {\bf 08} (2007) 059
[\href{http://www.arXiv.org/abs/0707.1038}{{\tt 0707.1038}}].

\bibitem{Douglas:2008jx}
M.~R. Douglas and G.~Torroba,  {\em {Kinetic terms in warped
  compactifications}}, JHEP {\bf 05} (2009) 013
[\href{http://www.arXiv.org/abs/0805.3700}{{\tt 0805.3700}}].

\bibitem{Shiu:2008ry}
G.~Shiu, G.~Torroba, B.~Underwood and M.~R. Douglas,  {\em {Dynamics of Warped
  Flux Compactifications}}, JHEP {\bf 06} (2008) 024
[\href{http://www.arXiv.org/abs/0803.3068}{{\tt 0803.3068}}].

\bibitem{Frey:2008xw}
A.~R. Frey, G.~Torroba, B.~Underwood and M.~R. Douglas,  {\em {The Universal
  Kaehler Modulus in Warped Compactifications}}, JHEP {\bf 01} (2009) 036
[\href{http://www.arXiv.org/abs/0810.5768}{{\tt 0810.5768}}].

\bibitem{Marchesano:2008rg}
F.~Marchesano, P.~McGuirk and G.~Shiu,  {\em {Open String Wavefunctions in
  Warped Compactifications}}, JHEP {\bf 04} (2009) 095
[\href{http://www.arXiv.org/abs/0812.2247}{{\tt 0812.2247}}].

\bibitem{Martucci:2009sf}
L.~Martucci,  {\em {On moduli and effective theory of N=1 warped flux
  compactifications}}, JHEP {\bf 05} (2009) 027
[\href{http://www.arXiv.org/abs/0902.4031}{{\tt 0902.4031}}].

\bibitem{Chen:2009zi}
H.-Y. Chen, Y.~Nakayama and G.~Shiu,  {\em {On D3-brane Dynamics at Strong
  Warping}}, Int. J. Mod. Phys. {\bf A25} (2010) 2493--2513
[\href{http://www.arXiv.org/abs/0905.4463}{{\tt 0905.4463}}].

\bibitem{Douglas:2009zn}
M.~R. Douglas,  {\em {Effective potential and warp factor dynamics}}, JHEP {\bf
  03} (2010) 071
[\href{http://www.arXiv.org/abs/0911.3378}{{\tt 0911.3378}}].

\bibitem{Kachru:2002sk}
S.~Kachru, M.~B. Schulz, P.~K. Tripathy and S.~P. Trivedi,  {\em {New
  supersymmetric string compactifications}}, JHEP {\bf 03} (2003) 061
[\href{http://www.arXiv.org/abs/hep-th/0211182}{{\tt hep-th/0211182}}].

\bibitem{Gurrieri:2002wz}
S.~Gurrieri, J.~Louis, A.~Micu and D.~Waldram,  {\em {Mirror symmetry in
  generalized Calabi-Yau compactifications}}, Nucl. Phys. {\bf B654} (2003)
  61--113
[\href{http://www.arXiv.org/abs/hep-th/0211102}{{\tt hep-th/0211102}}].

\bibitem{Lust:2008zd}
D.~Lust, F.~Marchesano, L.~Martucci and D.~Tsimpis,  {\em {Generalized
  non-supersymmetric flux vacua}}, JHEP {\bf 11} (2008) 021
[\href{http://www.arXiv.org/abs/0807.4540}{{\tt 0807.4540}}].

\bibitem{Banks:2010tj}
T.~Banks,  {\em {TASI Lectures on Holographic Space-Time, SUSY and
  Gravitational Effective Field Theory}},
\href{http://www.arXiv.org/abs/1007.4001}{{\tt 1007.4001}}.

\bibitem{Carroll:2009dn}
S.~M. Carroll, M.~C. Johnson and L.~Randall,  {\em {Dynamical compactification
  from de Sitter space}}, JHEP {\bf 11} (2009) 094
[\href{http://www.arXiv.org/abs/0904.3115}{{\tt 0904.3115}}].

\bibitem{Douglas:2010rt}
M.~R. Douglas and R.~Kallosh,  {\em {Compactification on negatively curved
  manifolds}}, JHEP {\bf 06} (2010) 004
[\href{http://www.arXiv.org/abs/1001.4008}{{\tt 1001.4008}}].

\bibitem{Koerber:2010bx}
P.~Koerber,  {\em {Lectures on Generalized Complex Geometry for Physicists}},
\href{http://www.arXiv.org/abs/1006.1536}{{\tt 1006.1536}}.

\bibitem{Danielsson:2009ff}
U.~H. Danielsson, S.~S. Haque, G.~Shiu and T.~Van~Riet,  {\em {Towards
  classical de Sitter solutions in string theory}}, JHEP {\bf 09} (2009) 114
[\href{http://www.arXiv.org/abs/0907.2041}{{\tt 0907.2041}}].

\bibitem{Janssen:1999sa}
B.~Janssen, P.~Meessen and T.~Ortin,  {\em {The D8-brane tied up: String and
  brane solutions in massive type IIA supergravity}}, Phys. Lett. {\bf B453}
  (1999) 229--236
[\href{http://www.arXiv.org/abs/hep-th/9901078}{{\tt hep-th/9901078}}].

\bibitem{Buscher:1987sk}
T.~H. Buscher,  {\em {A Symmetry of the String Background Field Equations}},
  Phys. Lett. {\bf B194} (1987)
59.

\bibitem{Hassan:1999mm}
S.~F. Hassan,  {\em {SO(d,d) transformations of Ramond-Ramond fields and space-
  time spinors}}, Nucl. Phys. {\bf B583} (2000) 431--453
[\href{http://www.arXiv.org/abs/hep-th/9912236}{{\tt hep-th/9912236}}].

\bibitem{Shelton:2005cf}
J.~Shelton, W.~Taylor and B.~Wecht,  {\em {Nongeometric Flux
  Compactifications}}, JHEP {\bf 10} (2005) 085
[\href{http://www.arXiv.org/abs/hep-th/0508133}{{\tt hep-th/0508133}}].

\bibitem{Douglas:2006es}
M.~R. Douglas and S.~Kachru,  {\em {Flux compactification}}, Rev. Mod. Phys.
  {\bf 79} (2007) 733--796
[\href{http://www.arXiv.org/abs/hep-th/0610102}{{\tt hep-th/0610102}}].

\bibitem{Silverstein:2004id}
E.~Silverstein,  {\em {TASI / PiTP / ISS lectures on moduli and microphysics}},
\href{http://www.arXiv.org/abs/hep-th/0405068}{{\tt hep-th/0405068}}.

\bibitem{Baumann:2010sx}
D.~Baumann, A.~Dymarsky, S.~Kachru, I.~R. Klebanov and L.~McAllister,  {\em
  {D3-brane Potentials from Fluxes in AdS/CFT}}, JHEP {\bf 06} (2010) 072
[\href{http://www.arXiv.org/abs/1001.5028}{{\tt 1001.5028}}].

\bibitem{progress2}
J.~Blaback, U.~Danielsson, D.~Junghans, T.~Van~Riet, T.~Wrase and M.~Zagermann.
\newblock {Work in progress}.

\bibitem{DeWolfe:2005uu}
O.~DeWolfe, A.~Giryavets, S.~Kachru and W.~Taylor,  {\em {Type IIA moduli
  stabilization}}, JHEP {\bf 07} (2005) 066
[\href{http://www.arXiv.org/abs/hep-th/0505160}{{\tt hep-th/0505160}}].

\bibitem{Bena:2009xk}
I.~Bena, M.~Grana and N.~Halmagyi,  {\em {On the Existence of Meta-stable Vacua
  in Klebanov- Strassler}},
\href{http://www.arXiv.org/abs/0912.3519}{{\tt 0912.3519}}.

\end{thebibliography}\endgroup
\bibliographystyle{utphysmodb}
\end{document}